# Mobile-D: An Agile Approach for Mobile Application Development

Pekka Abrahamsson[1], Antti Hanhineva[1], Hanna Hulkko[1], Tuomas Ihme[1], Juho Jäälinoja[1],
Mikko Korkala[2], Juha Koskela[1], Pekka Kyllönen[1], and Outi Salo[1]

[1]VTT Technical Research Centre of Finland, P.O. Box 1100, FIN-90571 Oulu
{firstname.lastname}@vtt.fi

[2]University of Oulu, Department of Information Processing Science, P.O. Box 3000, FIN-90014 Oulu
Mikko.Korkala@oulu.fi

## ABSTRACT

Mobile phones have been closed environments until recent years. The change brought by open platform technologies such as the Symbian operating system and Java technologies has opened up a significant business opportunity for anyone to develop application software such as games for mobile terminals. However, developing mobile applications is currently a challenging task due to the specific demands and technical constraints of mobile development. Furthermore, at the moment very little is known about the suitability of the different development processes for mobile application development. Due to these issues, we have developed an agile development approach called Mobile-D. The Mobile-D approach is briefly outlined here and the experiences gained from four case studies are discussed.

## Categories and Subject Descriptors
D.2.9 [**Software engineering**]: Management – *Life cycle, Productivity, Programming teams, Software process models*

## 1. INTRODUCTION
The mobile telecommunications industry has shown to be a highly competitive, uncertain and dynamic environment [1]. While so far mobile commerce applications have not been very successful, telecommunications companies are expecting a change in the near future due to the adoption of 3G technologies. This is believed to lead to a widespread adoption of mobile services in combination with mobile commerce applications [2]. Examples of such commercialized applications may include user- and location-specific mobile advertising, location-based services, and mobile financial services. The potential number of different mobile applications is virtually unlimited [3].

While mobile phones have been closed environments until recent years and their software them has been developed and maintained by the mobile terminal manufacturers themselves, the change brought by open platform technologies such as the Symbian operating system and Java-technologies has completely changed the situation. Now, basically anyone with the needed skills can develop applications for mobile terminals. This requires, however, a good knowledge of the specific characteristics and challenges of developing software for mobile devices.

Agile software development solutions can be seen to provide a good fit for the development of mobile applications [4]. However, the characteristics of mobile phones, terminals and the networking environment put some constraints on applying any of the existing agile software development methods. New, mobile specific approaches are needed. The Mobile-D approach presented in this work serves this purpose.

## 2. DEVELOPMENT OF MOBILE APPLICATIONS
Developing mobile applications is currently a very challenging task due to the specific demands and technical constraints of the mobile environment [5], such as

- limited capabilities and rapid evolution of terminal devices,
- various standards, protocols and network technologies,
- need to operate on a variety of different platforms,
- specific needs of mobile terminal users, and
- strict time to market requirements.

The principal differences in mobile devices are concerned with their physical characteristics, such as size, weight, display size, data input mechanism, and expandability. The technical characteristics of these devices – including processing power, memory space, battery capabilities and the operating system – also play an important role. The specific demands and characteristics of selected target devices need to be carefully considered in application development.

The product has to be of high quality from the start in order to function properly in dozens of different variations of existing and upcoming mobile phones. Furthermore, to be able to release a product even one week faster than the competitors can potentially lead to major international success, in view of the fact that the best selling mobile products such as downloadable java games costing from $3 to $5 are downloaded around the world thousands of times a week.

## 3. THE MOBILE-D APPROACH
To overcome the challenges involved in mobile application development, we have developed an agile development approach called the Mobile-D. The approach is based on Extreme Programming (development practices), Crystal methodologies (method scalability), and Rational Unified Process (life-cycle coverage).

The Mobile-D approach is optimized for a team of less than ten developers working in a co-located office space aiming at delivering a fully functional mobile application in a short time frame (i.e., less than 10 weeks). Mobile-D has been developed in co-operation with three companies developing mobile software products and services. The approach has been successfully assessed against the CMMI level 2 certification.

A development project, following the Mobile-D approach, is divided into five iterations. These phases are: set-up, core, core2, stabilize, and wrap-up. Each phase consists of three different types of development days: Planning Day, Working Day, and Release Day. If multiple teams are concurrently developing different parts of the same product, an Integration Day is also needed. The practices of the different phases comprise nine principal elements. These elements are as follows:

1. Phasing and Pacing
2. Architecture Line
3. Mobile Test-Driven Development
4. Continuous Integration
5. Pair Programming
6. Metrics
7. Agile Software Process Improvement
8. Off-Site Customer
9. User-Centered Focus

Most of these elements are well-known agile practices and have been specialized for mobile software development. The Architecture Line practice is an example of a Mobile-D specific element. The architecture line of an organization systematically captures current architectural knowledge about the patterns and solutions that have proven to be useful and working in the projects of the organization or in similar applications outside the organization. These typical software architectures, whose rationale is documented with pre-applied architectural patterns and working examples can be agilely utilized in Mobile-D projects.

The agile architecture line of the projects integrates the systematic piecemeal growth of software architecture with the phasing and pacing rhythm of the Mobile-D approach. The typical software architectures and patterns recorded in the architectural documentation of the projects provide not only trusted and high quality architectural solutions but also a sound rationale for these solutions.

To ensure proper operation of the product in multiple mobile platforms as early as possible, to enhance software design, and to improve software changeability, the development is required to be testing oriented. The Mobile Test-Driven Development (MTTD) practice employed in Mobile-D involves writing tests before actual implementation, automating unit testing procedures, and acceptance testing all features with the customer.

## 4. APPLYING MOBILE-D

The Mobile-D approach has been empirically tested and further developed in four case studies within the ENERGI (Industry-Driven Experimental Software Engineering Initiative) laboratory at VTT, The Technical Research Centre of Finland. These cases were concerned with new mobile phone extensions of database systems.

All of the applications from the case projects were delivered to market for real customers within 8 to10 weeks of calendar time. The developers in these projects consisted of both university students and software engineering professionals. Our goal is to further improve the Mobile-D approach within future ENERGI projects and also in industry.

When applying the Mobile-D approach in practice the following positive observations have been made: increased progress visibility, early identification and solving of technical problems (such as the difficulties with the J2ME and Symbian platforms), shared responsibility, efficient information sharing, high process-practice coherence, low defect density in released products, and constant development rhythm.

On the other hand, the following identified challenges will guide our further research efforts:

- High requirements for platform-specific development skills
- Variability and portability problems arise due to the diversity of characteristics in different mobile terminals.
- Weak tool support for Test-Driven Development in mobile environment.
- Adapting to rapid release cycles (management, customers)
- Need of highly disciplined practices (process fidelity)
- Attempts to use agile methods out-of-box lead to problems

## 5. ACKNOWLEDGEMENTS

The results presented here derive from the ICAROS project funded by The National Technology Agency of Finland (TEKES). The authors would like to thank the three participating companies (Qprojects Ltd, Sumea Interactive and BonumIT) for their active support in the development of the approach.